\documentclass[11pt,aps,prd,nofootinbib,preprint]{revtex4}
\usepackage{amsfonts}
\usepackage{mathrsfs}
\usepackage{graphicx}
\usepackage{amsmath}
\usepackage{amssymb}
\usepackage{subfigure}
\usepackage{epsfig}
\usepackage{graphicx}
\usepackage{color}
\newcommand{\bqa}{\begin{eqnarray}}
\newcommand{\eqa}{\end{eqnarray}}
\newcommand{\beq}{\begin{equation}}
\newcommand{\eeq}{\end{equation}}
\allowdisplaybreaks[1]
\graphicspath{{fig/}{dia/}} \DeclareGraphicsExtensions{.eps}
\begin{document}
	\title{Gluonic nature of the newly observed state $X(2600)$}
	
	\author{Sheng-Qi Zhang$^1$, Bing-Dong Wan$^{1,2}$, Liang Tang$^{3} $ and Cong-Feng Qiao$^{1}$\footnote{qiaocf@ucas.ac.cn}\vspace{+3pt}}
	
	\affiliation{$^1$ School of Physics, University of Chinese Academy of Science, Yuquan Road 19A, Beijing 100049 \\
		$^2$ School of Fundamental Physics and Mathematical Sciences, Hangzhou Institute for Advanced Study, UCAS, Hangzhou 310024, China\\
		$^3$ College of Physics and Hebei Key Laboratory of Photophysics Research and Application,
		Hebei Normal University, Shijiazhuang 050024, China
	}
	
	\author{~\\~\\}

	\begin{abstract}
	\vspace{0.3cm}
	Motivated by the newly observed resonance $X(2600)$ by BESIII Collaboration, we examine the trigluon glueball interpretation for it in the framework of  QCD sum rules. We evaluate the mass spectra of  the trigluon glueballs with quantum numbers $0^{-+}$ and $2^{-+}$ up to dimension 8 condensate in the operator product expansion. Our numerical results indicate that the mass of  the $2^{-+}$ trigluon glueball is about $2.66\pm 0.06$ GeV, which is consistent with the mass of the X(2600) within the uncertainties, while $0^{-+}$ has a mass of $2.01\pm0.14$ GeV. The possible decay channels of the $2^{-+}$ state are analyzed, which are crucial in decoding $X(2600)$'s internal structure and are hopefully measurable in BESIII, BEllEII, PANDA, and LHCb experiments.
	\end{abstract}
	 \maketitle
	\newpage

	\section{Introduction}
	Quantum chromodynamics (QCD) as the fundamental theory of hadronic interaction has been accurately tested up to the 1\% level in the high energy region because of asymptotic freedom \cite{Gross:1973id, Politzer:1973fx}. However, the nonperturbative solution of hadron spectrum is hard to be derived from first principles due to the QCD confinement \cite{Wilson:1974sk}. A unique attempt to improve our understanding of the nonperturbative aspects of QCD is to study glueballs, bound states of pure gluons, in which the gauge field plays a more important dynamical role than in the conventional hadrons, which has created much interest in theory and experiment for a long time.
		
	The glueballs have been searched experimentally in the last fifty years, and the properties of such states were explored in depth in a variety of theories, such as lattice QCD \cite{Meyer:2004gx, Athenodorou:2020ani, Morningstar:1999rf, Chen:2005mg,  Gregory:2012hu}, the flux tube model \cite{Isgur:1984bm, Faddeev:2003aw}, the MIT bag model \cite{Jaffe:1975fd, Chodos:1974je, Carlson:1982er, Donoghue:1980hw}, the constituent gluon model \cite{Szczepaniak:1995cw, Llanes-Estrada:2005bii}, the AdS/QCD model \cite{deTeramond:2005su, Brunner:2018wbv,Chen:2015zhh,Zhang:2021itx}, and QCD sum rules \cite{Shifman:1978by, Novikov:1979va, Shuryak:1982dp, Narison:1988ts, Huang:1998wj, Forkel:2000fd, Hao:2005hu, Qiao:2014vva, Tang:2015twt, Narison:2021xhc, Chen:2021bck, Chen:2021cjr, Bagan:1990sy}. By means of these techniques, the two-gluon glueballs have been studied extensively. Many calculations tell of the $ 0^{++} $ scalar glueball having a mass of about $ 1.6\pm0.3 $ GeV \cite{Meyer:2004gx, Athenodorou:2020ani, Morningstar:1999rf, Chen:2005mg, Gregory:2012hu, Isgur:1984bm, Carlson:1982er, Szczepaniak:1995cw, Chen:2015zhh, Novikov:1979va, Shuryak:1982dp, Narison:1988ts, Huang:1998wj, Forkel:2000fd, Bagan:1990sy, Chen:2021bck}, which suggests that the scalar mesons $f_0(1370)$, $ f_0(1500)$, and $ f_0(1710) $ \cite{BES:2004twe, Gray:1983cw, Serpukhov-Brussels-AnnecyLAPP:1983xdr, Burke:1982am, Etkin:1982se, Edwards:1981ex} may be the possible candidates of glueballs or glue rich objects. Cheng \textit{et al.} deduced a $ 0^{-+} $ pseudoscalar glueball, if exists, may have a mass around 1400 MeV from an $ \eta$-$\eta^{\prime}$-G mixing formalism \cite{Cheng:2008ss}, which is consistent with the viewpoint in Refs.~\cite{Chanowitz:1980gu, Donoghue:1980hw, Ishikawa:1980xv, Li:2007ky, Faddeev:2003aw, Close:1996yc}. Although the lattice QCD predictions indicate that the two-gluon pseudoscalar glueball lies above 2.2 GeV \cite{Meyer:2004gx, Athenodorou:2020ani, Morningstar:1999rf, Chen:2005mg,  Gregory:2012hu}, the pseudoscalar state $ \eta(1405) $ observed in $ \eta\pi\pi $ channel of $ J/\psi $ decay \cite{BESIII:2011nqb} seems to contain a large gluon component \cite{Wan:2019fuk}. As for the $ 2^{-+} $ pseudotensor glueball, Carlson \textit{et al.} proposed a two-gluon pseudotensor glueball with mass about 1.9 GeV by exploiting MIT bag model \cite{Carlson:1982er}.

\begin{figure}
	 	\includegraphics[width=.40\textwidth]{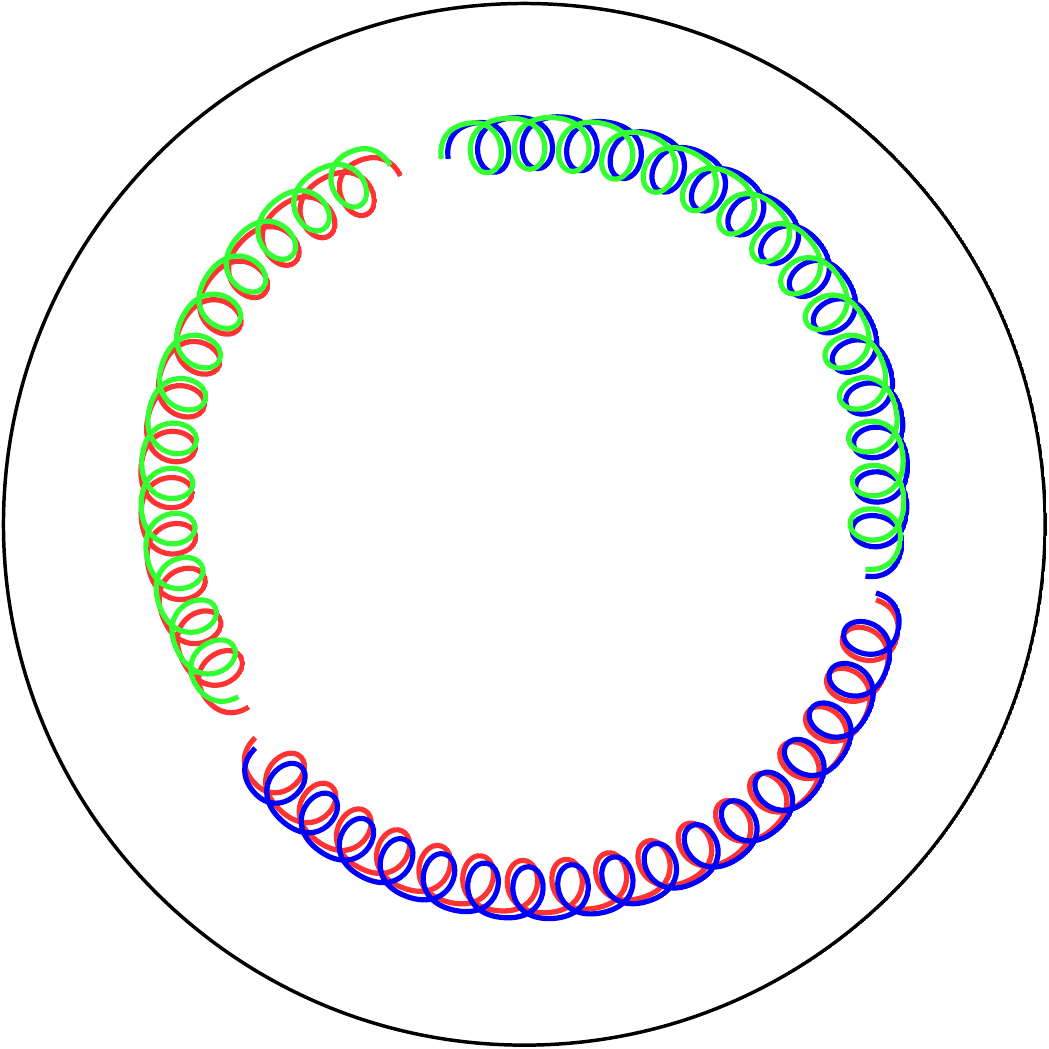}
	 	\caption{  A sketch of the trigluon glueball. Each gluon line represents a dynamic gluon.}
	 	\label{cartoon}
	 \end{figure}

	The studies about the glueball mentioned above are mostly limited to the two-gluon glueball, the leading Fock state in Fock space expansion \cite{Bagan:1990sy, Narison:1988ts, Hao:2005hu}. It should be noted that glueballs constructed by three dynamic gluons, trigluon glueball, can possess every possible quantum number, as shown in Fig.~\ref{cartoon}. Although the two-gluon glueball and trigluons glueball are both permitted by QCD, there is still no conclusive experimental evidence for their existence, which is partly due to the lack of necessary knowledge of their production and decay properties and the strong expected mixing between glueballs and quark states \cite{Chen:2022asf}. Having a great collection of $J/\psi$ data, BESIII Collaboration carefully examined the $J/\psi \rightarrow \pi^+\pi^-\eta^\prime$ process \cite{BES:2005ega,BESIII:2010gmv,BESIII:2016fbr,BESIII:2021xoh,BESIII:2020kpr,BESIII:2019sfz,BESIII:2017hup} which is one of most favorite modes to search for pseudoscalar glueball \cite{Amsler:2004ps,Eshraim:2012jv}.
	
	 Recently, by scrutinizing the radiative decays of $J/\psi$, BESIII Collaboration observed a structure in $\pi^+ \pi^- \eta^\prime $ invariant mass of about $2.62$ GeV with significance greater than $20~\sigma$ \cite{BESIIICollaboration:2022qjc}. From its known decay products, the new state most likely possesses quantum number of $0^{-+}$ or $2^{-+}$. Since the intermediate resonance $f_0(1500)$ and final state $\eta^\prime$ are both glueball candidates, or at least contain a large content of gluons, it is reasonable to consider $X(2600)$ as a glue-rich object or a glueball.
	
	In this work, to understand the gluonic nature of the $ X(2600) $, we investigate the trigluon glueball state with quantum numbers of $ 0^{-+} $ and $ 2^{-+} $ by virtue of the QCD sum rules developed more than 40 years ago by Shifman, Vainshtein, and Zakharov (SVZ) \cite{Shifman:1978by}. The QCD sum rules (QCDSR) has some peculiar advantages in studying hadron spectrum involving nonperturbative effect of QCD. Rather than a phenomenological model, QCDSR is a QCD based theoretical framework incorporating nonperturbative effects universally order by order, which has already achieved a lot in the study of hadron spectroscopy and decays. In order to evaluate the mass spectra of the glueballs, one has to construct the appropriate currents that possesses the foremost information about the concerned hadron, such as quantum numbers and the constituent quarks or gluons. By exploiting the current, the two-point correlation function can be constructed, which has two representations: the QCD representation and the phenomenological representation. The QCDSR will be formally established after equating these two representations, from which the mass and decay width of hadron can be obtained.
	
	The rest of the paper is organized as follows. In Sec.~\ref{Formalism}, we construct corresponding currents for the trigluon glueball and present the basic formulas in our evaluation. In Sec.~\ref{Numerical}, we give the analytical and numerical results for each possible trigluon glueball. In Sec.~\ref{Decay}, a brief discussion on possible bigluon- and trigluon-glueball decay modes are give. The last section is left for conclusions.
	
	\section{Formalism}\label{Formalism}
	
	To calculate the mass spectra of $0^{-+} $ and $2^{-+}$ trigluon glueballs, the appropriate interpolating currents have to be established. Although there are a number of currents that satisfy these quantum numbers, only a limited number of currents remain after taking into account the constrains of gauge invariance, Lorentz invariance, and $\text{SU}_c(3)$ symmetry. For the $0^{-+}$ trigluon glueball, they are
	\begin{eqnarray}
		j^{0^{-+}, \; A}(x)&\,=\,& g_s^3 \,f^{a b c} \,\tilde{G}^a_{\mu \nu}(x)\, G^b_{\nu \rho}(x)\, G^c_{\rho \mu}(x)\, , \label{current-0-+A}\\
		j^{0^{-+}, \; B}(x)&\,=\,& g_s^3 \,f^{a b c} \,G^a_{\mu \nu}(x)\, \tilde{G}^b_{\nu \rho}(x)\, G^c_{\rho \mu}(x)\, , \label{current-0-+B}\\		
		j^{0^{-+}, \; C}(x)&\,=\,& g_s^3 \,f^{a b c} \,G^a_{\mu \nu}(x)\, G^b_{\nu \rho}(x)\, \tilde{G}^c_{\rho \mu}(x)\, , \label{current-0-+C}\\
		j^{0^{-+}, \; D}(x)&\,=\,& g_s^3 \,f^{a b c} \,\tilde{G}^a_{\mu \nu}(x)\, \tilde{G}^b_{\nu \rho}(x)\, \tilde{G}^c_{\rho \mu}(x)\, . \label{current-0-+D}
	\end{eqnarray}
    For the $2^{-+}$ trigluon glueball, they are
	\begin{eqnarray}
		j_{\mu\sigma}^{2^{-+}, \; A}(x)&\,=\,& g_s^3 \,f^{a b c} \,\tilde{G}^a_{\mu \nu}(x)\, G^b_{\nu \rho}(x)\, G^c_{\rho \sigma}(x)\, , \label{current-2-+A}\\
		j_{\mu\sigma}^{2^{-+}, \; B}(x)&\,=\,& g_s^3 \,f^{a b c} \,G^a_{\mu \nu}(x)\, \tilde{G}^b_{\nu \rho}(x)\, G^c_{\rho \sigma}(x)\, , \label{current-2-+B}\\
		j_{\mu\sigma}^{2^{-+}, \; C}(x)&\,=\,& g_s^3 \,f^{a b c} \,G^a_{\mu \nu}(x)\, G^b_{\nu \rho}(x)\, \tilde{G}^c_{\rho \sigma}(x)\, . \label{current-2-+C}\\
		j_{\mu\sigma}^{2^{-+}, \; D}(x)&\,=\,& g_s^3 \,f^{a b c} \,\tilde{G}^a_{\mu \nu}(x)\, \tilde{G}^b_{\nu \rho}(x)\, \tilde{G}^c_{\rho \sigma}(x)\, . \label{current-2-+D}
	\end{eqnarray}
    Here, $a$, $b$, and $c$ are color indices, $\mu$, $\nu$, $\rho$, and $\sigma$ denote Lorentz indices, $f^{a b c}$ stands for totally antisymmetric $\text{SU}_c(3)$ structure constant, $G^a_{\mu \nu}$ represents the gluon field strength tensor, and $\tilde{G}^a_{\mu \nu}$ is the dual gluon field strength tensor defined as $\tilde{G}^a_{\mu \nu} = \frac{1}{2} \epsilon_{\mu \nu \kappa \tau} G^a_{\kappa \tau}$. Moreover, it should be noted that current (\ref{current-0-+B}) and (\ref{current-0-+C}) can be proved to be equivalent to current (\ref{current-0-+A}), and current (\ref{current-2-+C}) is equivalent to current (\ref{current-2-+A}). Thus, in our calculation, we only concentrate on currents (\ref{current-0-+A}), (\ref{current-0-+D}), (\ref{current-2-+A}), (\ref{current-2-+B}), and (\ref{current-2-+D}), which are not equivalent.

    Equipped with interpolating currents, the two-point correlation functions can be readily established, i.e.,
	\begin{eqnarray}
		\Pi^k(q^2) &=& i \int d^4 x e^{i q \cdot x} \langle 0 | T \{ j^{k} (x),\;  j^{k} (0) \} |0 \rangle\;,\label{pi1}\\
		\Pi^k_{\mu\nu, \rho\sigma}(q^2) &=& i \int d^4 x e^{i q \cdot x} \langle 0 | T \{ j^k_{\mu\nu} (x),\;  j_{\rho\sigma}^k (0) \} |0 \rangle\;, \;\label{pi2}
	\end{eqnarray}
	where $j (x)$ and $j_{\mu\nu} (x)$ stand for the interpolating currents coupled to the trigluon glueball with $J= $ 0 and 2, respectively, and $|0\rangle$ denotes the physical vacuum. The superscript $k$ represents $A $ and $ D$ for the $0^{-+}$ trigluon glueball, while $A $, $ B$, and $ D$ for the $2^{-+}$ trigluon glueball.
	
	Eq.~(\ref{pi2}) has the following Lorentz structure \cite{Qiao:2013dda,Chen:2011qu}
	\begin{eqnarray}
		i \int d^4 x e^{i q \cdot x} \langle 0 | T \{ j^k_{\mu\nu} (x),\;  j_{\rho\sigma}^k (0) \} |0 \rangle = T_{\mu \nu, \, \rho
				\sigma}\,\,\Pi^k_{2^{-+}}(q^2)+\cdots \; ,
	\end{eqnarray}
	 where “$\cdots$” stands for other Lorentz structures independent of the correlation function $\Pi^k_{2^{-+}}(q^2) $, and $T_{\mu \nu, \,
	 		\rho \sigma}$ is the unique Lorentz tensor of the fourth rank
	 	constructed from $g_{\mu \nu}$ and $q_\mu$:
	\begin{eqnarray}
		T_{\mu \nu, \, \rho \sigma}=\frac{1}{2} \big[g^t_{\mu \rho}(q)
		g^t_{\nu \sigma}(q) + g^t_{\mu \sigma}(q) g^t_{\nu \rho}(q) -
		\frac{2}{3} g^t_{\mu \nu}(q) g^t_{\rho \sigma}(q) \big] \;,
	\end{eqnarray}
	which satisfies the following properties:
	\begin{eqnarray}
		& & T_{\mu \nu, \, \rho \sigma} = T_{\rho \sigma, \, \mu \nu}, \;
		\; q^\mu
		T_{\mu \nu, \, \rho \sigma}  = 0 \;, \nonumber \\
		& & g_t^{\mu \nu}(q) T_{\mu \nu, \, \rho \sigma} = 0 \; .
	\end{eqnarray}
	Here, $g^{t}_{\mu\nu}(q)=g_{\mu\nu}-q_\mu q_\nu/q^2 $ with $g_{\mu\nu} $ the Lorentz metric tensor.
	
	The correlation function on the QCD representation can be obtained by the operator product expansion (OPE),
	\begin{eqnarray}
		\mathrm{\Pi}_{J^{PC} }^{k, \; \mathrm{QCD}}(q^2) & \!\! = \!\! & \left(a_0 + a_1 \ln\frac{-q^2}{\mu^2}\right) (q^2)^4 + \left( b_0 + b_1 \ln \frac{-q^2}{\mu^2} \right)(q^2)^{2} \langle \alpha_s G^2 \rangle  \nonumber \\
		& \!\!+ \!\! & \left( c_0 + c_1 \ln\frac{- q^2}{\mu^2} \right) (q^2) \,\langle g_s^3 G^3 \rangle + \left(d_0+ d_1 \ln\frac{-q^2}{\mu^2}\right) \langle \alpha_s^2 G^4 \rangle \;, \label{correlation-function-QCD}
	\end{eqnarray}
	where $\langle \alpha_s G^2 \rangle$, $\langle g_s^3 G^3\rangle$, and $\langle \alpha_s^2 G^4 \rangle$ represent condensates with different dimensions; $\mu$ is the renormalization scale. Here, we use $ a_0$, $ a_1$, $ b_0$, $ b_1$, $ c_0$, $ c_1$, $d_0$, and $d_1$ to represent Wilson coefficients of operators with different dimensions.
	
	On the phenomenological representation, adopting the pole plus continuum parametrization of the hadronic spectral density, the imaginary part of the correlation function can be described as:
	\begin{eqnarray}
	 \frac{1}{\pi} \text{Im}\,\Pi_{J^{PC}}^{k,\,\text{phe}}& = &(\lambda^{k}_{J^{PC}})^2\, \delta \left(s - (M_{J^{PC}}^{k})^2 \right)  +  \rho_{J^{PC}}^{k,\,\text{cont}} (s) \theta(s - s_0) \;\label{correlation-function-PHE} .
	\end{eqnarray}
	Here, $\lambda^{k}_{J^{PC}} $ is the coupling constant; $M_{J^{PC}}^{k} $ stands for the mass of the $ J^{PC}$ trigluon glueball. $\rho_{J^{PC}}^{k,\,\text{cont}} (s) $ represents the spectral density which contains continuous spectrum and high excited states above the continuum threshold $\sqrt{s_0} $.
	
	Employing the dispersion relation on both QCD and phenomenological representation, i.e.,
	\begin{eqnarray}\label{qcd-dis}
		\Pi_{J^{PC}}^{k}(q^2) &\,= \,& \frac{1}{\pi} \int_0^\infty  \frac{\text{Im}\,  \Pi_{J^{PC}}^{k}(s) }{s - q^2} \,ds +\cdots  ,
	\end{eqnarray}
    where `$\cdots$' represents the subtraction terms, then one can establish connection between the QCD side and the phenomenological side,
	\begin{eqnarray}\label{main}
		 \frac{1}{\pi} \int_0^\infty  \frac{\text{Im}\,  \Pi_{J^{PC}}^{k,\,\text{QCD}}(s) }{s - q^2} \,ds+ \cdots&=& \frac{(\lambda^k_{J^{PC}})^2}{(M^k_{J^{PC}})^2 - q^2} + \int_{s_0}^\infty  \frac{\rho^{k,\,\text{cont}}_{J^{PC}}(s)\,\theta(s-s_0)}{s - q^2}\,ds \;.
	\end{eqnarray}
	
	To limit the contributions of higher order condensates on the QCD side and the contributions of high excited and continuum states in the phenomenological side, an effective way is to impose Borel transformation simultaneously on both sides. The form of Borel transformation is as follows,
	\begin{equation}\label{borel}	
		{\cal B}\! \left[ g(q^2) \right] \equiv g(\tau) =
		\lim\limits_{\tiny \begin{matrix} -q^2, n \rightarrow \infty \\ \frac{-q^2}{n} = \frac{1}{\tau} \end{matrix}}
		\frac{(q^2)^{n}}{(n-1)!} \left( -\frac{\partial}{\partial q^2} \right)^n \!g(q^2)\;.
	\end{equation}
	Here, $g(q^2) $ is the function of the independent variable $ q^2$ and $ \tau $ is a free parameter which is called the Borel parameter.
	
	Performing the Borel transformation to Eq.~(\ref{main}) one can obtain,
	\begin{eqnarray}
		\label{borelope}
		 \frac{1}{\pi} \int_{0}^{\infty} \!  \: \text{Im}\, \Pi_{J^{PC}}^{k,\,\text{QCD}}(s)~e^{-s \tau}\,ds &=&(\lambda^k_{J^{PC}})^2 ~e^{-(M^k_{J^{PC}})^2 ~\tau} +
		\int_{s_0}^{\infty} \! \:\rho^{k,\,\text{cont}}_{J^{PC}}(s) ~e^{-s \tau}\,ds ~.	\end{eqnarray}
	Here, the subtraction terms in Eq.(\ref{qcd-dis}) turns to be zero after the Borel transformation.
	
	Adopting quark-hadron duality
		\begin{eqnarray}
		\frac{1}{\pi} \int_{s_0}^{\infty} \!  \: \text{Im}\, \Pi_{J^{PC}}^{k,\,\text{QCD}}(s)~e^{-s \tau}\,ds&\simeq&
		\int_{s_0}^{\infty} \!  \:\rho^{k,\,\text{cont}}_{J^{PC}}(s) ~e^{-s \tau}\,ds ~,
		\label{duality}
	\end{eqnarray}
	then one can obtain the main equation of QCDSR,
	\begin{eqnarray}
		(\lambda^k_{J^{PC}})^2 ~e^{-(M^k_{J^{PC}})^2 ~\tau} = \frac{1}{\pi} \int_{0}^{s_0} \!  \: \text{Im}\, \Pi_{J^{PC}}^{k,\,\text{QCD}}(s)~e^{-s \tau}\,ds\;,
	\end{eqnarray}
	from which the mass of  trigluon glueball with different $J^{PC} $ can be expressed as
	\begin{eqnarray}
		M^{k}_{J^{PC}}(\tau, s_0) = \sqrt{ \frac{L^k_{J^{PC}, \; 1} (\tau, s_0)}{L^k_{J^{PC}, \; 0} (\tau, s_0)}} \; , \label{mass}
	\end{eqnarray}
	where the moments $L^{k}_{J^{PC}, \; 0}$ and $L^{k}_{J^{PC}, \; 1}$ can be now reached,
	\begin{eqnarray}
		L^{k}_{J^{PC}, \; 0}(\tau, s_0) & = & \frac{1}{\pi}\, \int_{0}^{s_0} \!  \: \text{Im}\, \Pi_{J^{PC}}^{k,\,\text{QCD}}(s)~e^{-s \tau}\,ds  \; , \label{R0} \\
		L^{k}_{J^{PC}, \; 1}(\tau, s_0) & = &\frac{1}{\pi}\, \int_{0}^{s_0} \!  \: \text{Im}\, \Pi_{J^{PC}}^{k,\,\text{QCD}}(s)~s\,e^{-s \tau}\,ds \;. \label{R1}
	\end{eqnarray}
	Here, $L^{k}_{J^{PC}, \; 1}(\tau, s_0)$ is obtained via $L^{k}_{J^{PC}, \; 1}(\tau, s_0)= - \partial L^{k}_{J^{PC}, \; 0}(\tau, s_0)/ \partial \tau$.
	
	\section{Numerical analyses}\label{Numerical}
	\begin{figure}
	\includegraphics[width=.90\textwidth]{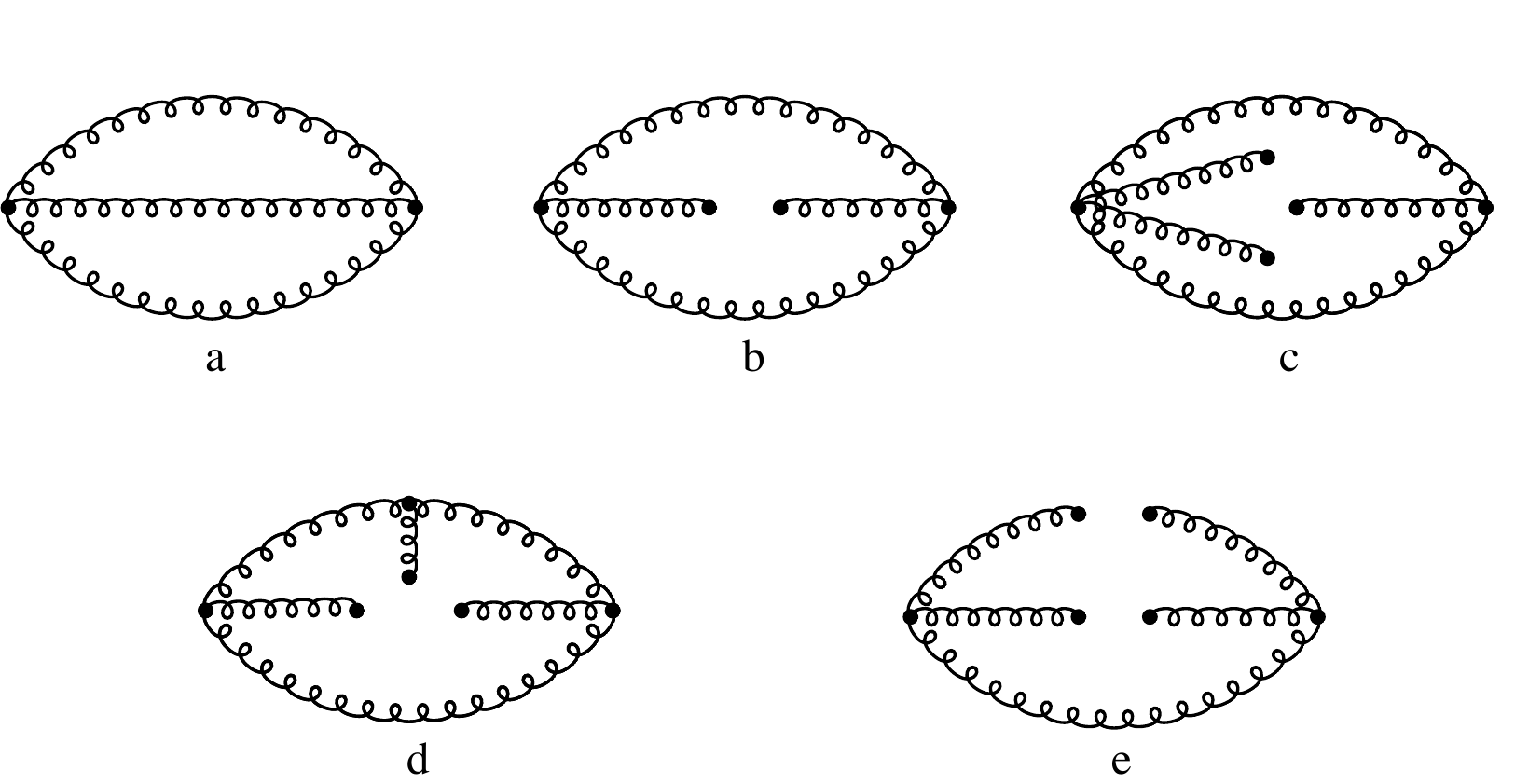}
	\caption{ The typical Feynman diagrams of trigluon glueball in the scheme of QCD sum rules. (a) Diagram for the perturbative term; (b) for the two-gluon condensate terms; (c) and (d) for the three-gluon condensate terms; (e) for the four-gluon condensate terms.}
	\label{feyn}
    \end{figure}
	The Feynman diagrams of the correlation function in our calculation are shown in Fig.~\ref{feyn}. After a lengthy calculation, the Wilson coefficients in Eq.~(\ref{correlation-function-QCD}) can be obtained. For the $0^{-+} $ trigluon glueball, they are
	\begin{eqnarray}
		\begin{aligned}
			& a_0^{A} \, = \,\frac{181}{400} \frac{\alpha_s^3}{\pi} \; ,& &a_1^{A} \,= \, -\frac{3}{10} \frac{\alpha_s^3}{\pi} \;,  & & b_0^{A} \, =\, 0 \;, & & b_1^{A} \, = \, 0 \; , \\
			& c_0^{A} \, = \,\frac{21}{4}\;  \alpha_s^2\; ,& &c_1^{A} \,= \, -\frac{27}{2} \;\alpha_s^2 \;,  & & d_0^{A} \, =\, 0 \;, & & d_1^{A} \, = \, 0 \; , \\
			& a_0^D \, = \,\frac{261}{400} \frac{\alpha_s^3}{\pi} \; ,& &a_1^D \,= \, -\frac{3}{10} \frac{\alpha_s^3}{\pi} \;,  & & b_0^D \, =\, 18\; \pi \alpha_s^2 \;, & & b_1^D \, = \, 0 \; , \\
			& c_0^D \, = \,\frac{189}{4}\; \alpha_s^2\; ,& &c_1^D \,= \, -\frac{27}{2} \;\alpha_s^2 \;,  & & d_0^D \, =\, -36\; \pi^3 \alpha_s \;, & & d_1^D \, = \, 0 \; ,\nonumber
		\end{aligned}
	\end{eqnarray}
	where the analytical results of case $D $ are directly obtained from Ref.~\cite{Hao:2005hu}. We notice that $a_1^k $, $b_1^k $, $c_1^k $, and $d_1^k $ are equal for case $A$ and $ D$, which implies that the mass spectrum of case $A$ is exactly the same as case $ D$ because it depends entirely on the imaginary part of the correlation function, i.e. Wilson coefficients $a_1^k $, $b_1^k $, $c_1^k $, and $d_1^k $. However, the complete correlation functions are different, which implies there may be two $0^{-+}$ trigluon glueballs with degeneracy.
	
	For the $2^{-+} $ trigluon glueball, the Wilson coefficients are
		\begin{eqnarray}
		\begin{aligned}
			& a_0^{A} \, = \,\frac{229}{2^6\times 3^2\times 7^2 \times 5} \frac{\alpha_s^3}{\pi} \; ,& &a_1^{A} \,= \, -\frac{1}{168} \frac{\alpha_s^3}{\pi} \;,  & & b_0^{A} \, =\, -\frac{32}{15}\; \pi \alpha_s^2 \;, & & b_1^{A} \, = \, 0 \; , \\
			& c_0^{A} \, = \,-\frac{1}{4}\, \alpha_s^2\; ,& &c_1^{A} \,= \, \;\frac{1}{4}\;\alpha_s^2 \;,  & & d_0^{A} \, =\, 0 \;, & & d_1^{A} \, = \, 0 \; , \\
			& a_0^B \, = \,\frac{54937}{2^6\times 3^2\times 5^2 \times 7^2} \frac{\alpha_s^3}{\pi} \; ,& &a_1^B \,= \, -\frac{3}{56} \frac{\alpha_s^3}{\pi} \;,  & & b_0^B \, =\, 0 \;, & & b_1^B \, = \, 0 \; , \\
			& c_0^B \, = \,\frac{383}{120}\;  \alpha_s^2\; ,& &c_1^B \,= \, \frac{9}{4} \;\alpha_s^2 \;,  & & d_0^B \, =\, 0 \;, & & d_1^B \, = \, 0 \; , \\
			& a_0^D \, = \,-\frac{1451}{2^6\times 3^2\times 7^2 \times 5} \frac{\alpha_s^3}{\pi} \; ,& &a_1^D \,= \, -\frac{1}{168} \frac{\alpha_s^3}{\pi} \;,  & & b_0^D \, =\, 0 \;, & & b_1^D \, = \, 0 \; , \\
			& c_0^D \, = \,-\frac{247}{192}\;  \alpha_s^2\; ,& &c_1^D \,= \,  \frac{1}{4}\;\alpha_s^2 \;,  & & d_0^D \, =\, 0\;, & & d_1^D \, = \, 0 \; , \nonumber
		\end{aligned}
	\end{eqnarray}
    where we notice that $ a_1^B=9\,a_1^A=9\,a_1^D$ and $ c_1^B=9\,c_1^A=9\,c_1^D$, which implies that the mass curves of cases $A$, $ B$, and $D$ will be exactly equal based on Eq.~(\ref{mass}). Similar to the cases of $0^{-+}$ trigluon glueball, the complete correlation functions of case $ A$, $ B$, and $D$ are different, which suggests there may exist three $ 2^{-+} $ trigluon glueballs with degeneracy.
	
	To perform numerical calculation, the following inputs are adopted \cite{Huang:1998wj, Forkel:2000fd, Bagan:1990sy, Ioffe:2005ym, Chen:2021cjr}:
		\begin{eqnarray}\label{input}
		\begin{aligned}
			&\langle \alpha_s G^2\rangle = (0.005\pm0.004)\times \pi \, \text{GeV}^4 \; ,\;
			\langle g_s^3 G^3\rangle =  \langle \alpha_s G^2\rangle \times (8.2\pm1.0) \, \text{GeV}^2 \; ,\\
			&\langle \alpha_s^2 G^4\rangle=\frac{9}{16} \langle \alpha_s G^2\rangle ^2\; , \;\Lambda_{\overline{\text{MS}}} = 300 \, \text{MeV} \; ,\;
			\alpha_s = \frac{-4\pi}{11 \ln (\tau\,\Lambda^2_{\overline{\text{MS}}})} \; .
		\end{aligned}
	\end{eqnarray}
	
	Moreover, in the calculation two additional free parameters, i.e., the continuum threshold $s_0$ and the Borel parameter $\tau$, are introduced in the establishment of QCDSR. In general, the values of $s_0$ and $\tau$ can be determined by two criteria \cite{Shifman:1978by, Reinders:1984sr, Colangelo:2000dp}. First, the pole contribution(PC). In order to extract the properties of ground-state hadrons, the pole contribution has to dominate the spectrum. As discussed in Refs.~\cite{Wan:2021vny, Wang:2016gxp, Tang:2021zti, Wan:2019ake}, the large power of $s$ in the spectrum suppresses the PC value. Thus, the pole contribution can be selected larger than $(40-60)\%$ \cite{Wan:2022xkx,Xin:2021wcr,Wang:2021qus} for the trigluon glueball. This kind of criterion can be formulated as follows:
	\begin{eqnarray}
		R^{k, \; \text{PC}}_{J^{PC}} &=& \frac{L^{k}_{J^{PC} , \; 0}(\tau,s_0)}{L^{k}_{J^{PC} , \; 0}(\tau,\infty)} \; . \label{ratio-PC}
	\end{eqnarray}
	
	The second criterion is the convergence of OPE, that is the condensate term with neglected power corrections remains a small fraction of all terms in the truncated OPE. In other words, the relative contribution of the lowest dimension of condensate needs to be less than $ 50\%$. Therefore, one needs to calculate the proportions of each condensate term in the QCD side. In our calculation, two gluons condensate $ \langle \alpha_s G^2\rangle$ do not contribute to the expansion of OPE because all $b^k_1 $ are zero. Thus, this criterion can be formulated as follows:
	\begin{eqnarray}
		R^{k, \; \langle\text{G}^3\rangle}_{J^{PC}} &=&\frac{\int_{0}^{s_0} \!  \: \text{Im}\, \Pi_{J^{PC}}^{k,\,\langle g_s^3 G^3\rangle}(s)~e^{-s \tau}\,ds}{\int_{0}^{s_0} \!  \: \text{Im}\, \Pi_{J^{PC}}^{k,\,\text{QCD}}(s)~e^{-s \tau}\,ds}  \; . \label{ratio-GGG}
	\end{eqnarray}
	
	In order to search a proper value of $\sqrt{s_0} $, we implement a similar analysis as in Refs.~\cite{Tang:2015twt,Wan:2020fsk, Wan:2020oxt, Yang:2020wkh}. That is, one needs to find an optimal window, in which the mass of the trigluon glueball is independent of the Borel parameter $\tau $ as much as possible. Since the continuum threshold $\sqrt{s_0} $ is correlated with the onset of excited states in the channel of the current $J^{PC} $ \cite{Colangelo:2000dp}, a number of $\sqrt{s_0} $ are taken into account. The central value of $\sqrt{s_0} $ can be obtained by the above procedure. In practice, we vary $\sqrt{s_0} $ by 0.1 $\text{GeV}$, which gives the lower and upper bounds and hence the uncertainties of $\sqrt{s_0} $.
	
\begin{figure}[ht]
		\begin{center}
			\includegraphics[width=6.8cm]{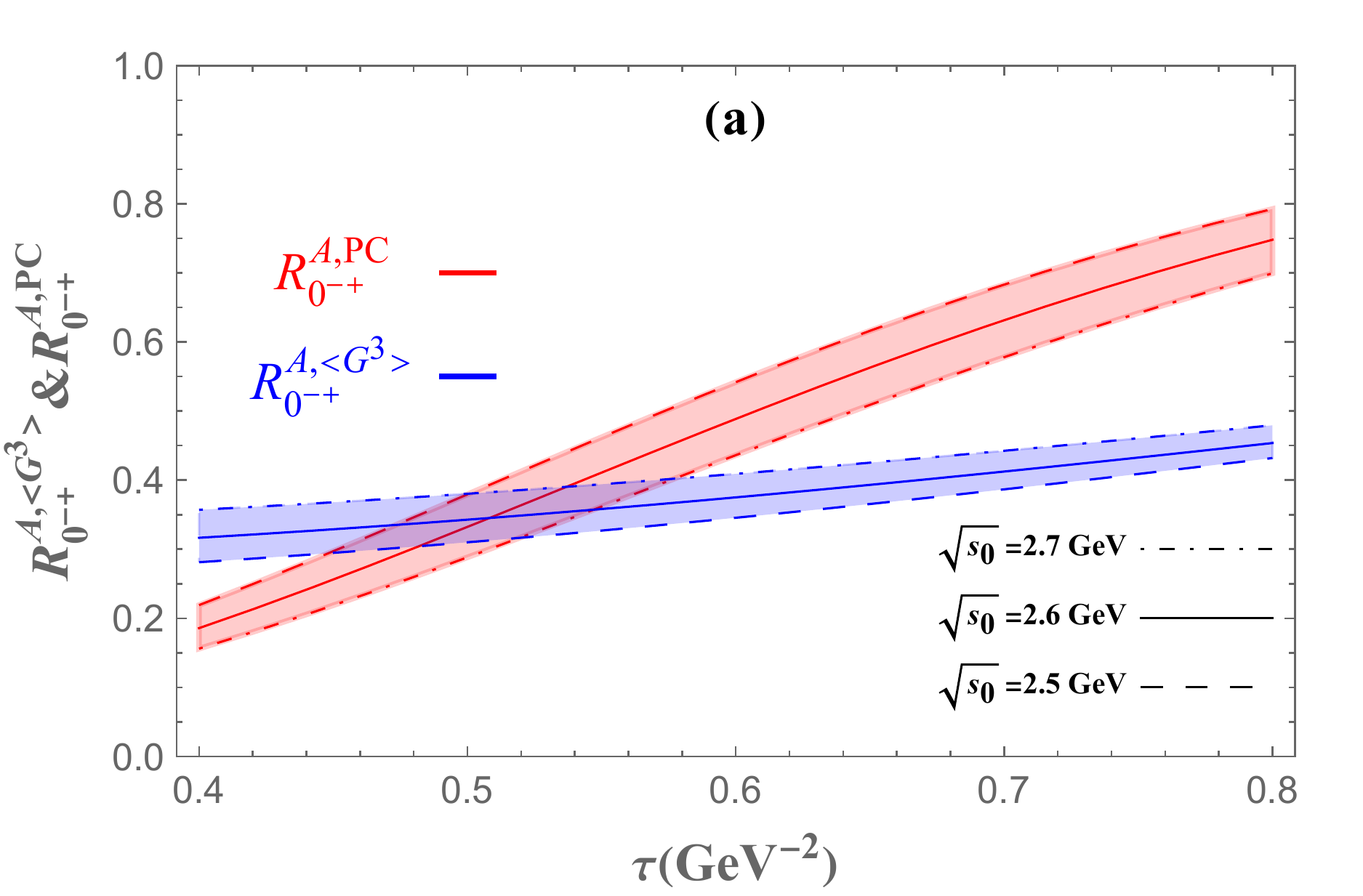}
			\includegraphics[width=6.8cm]{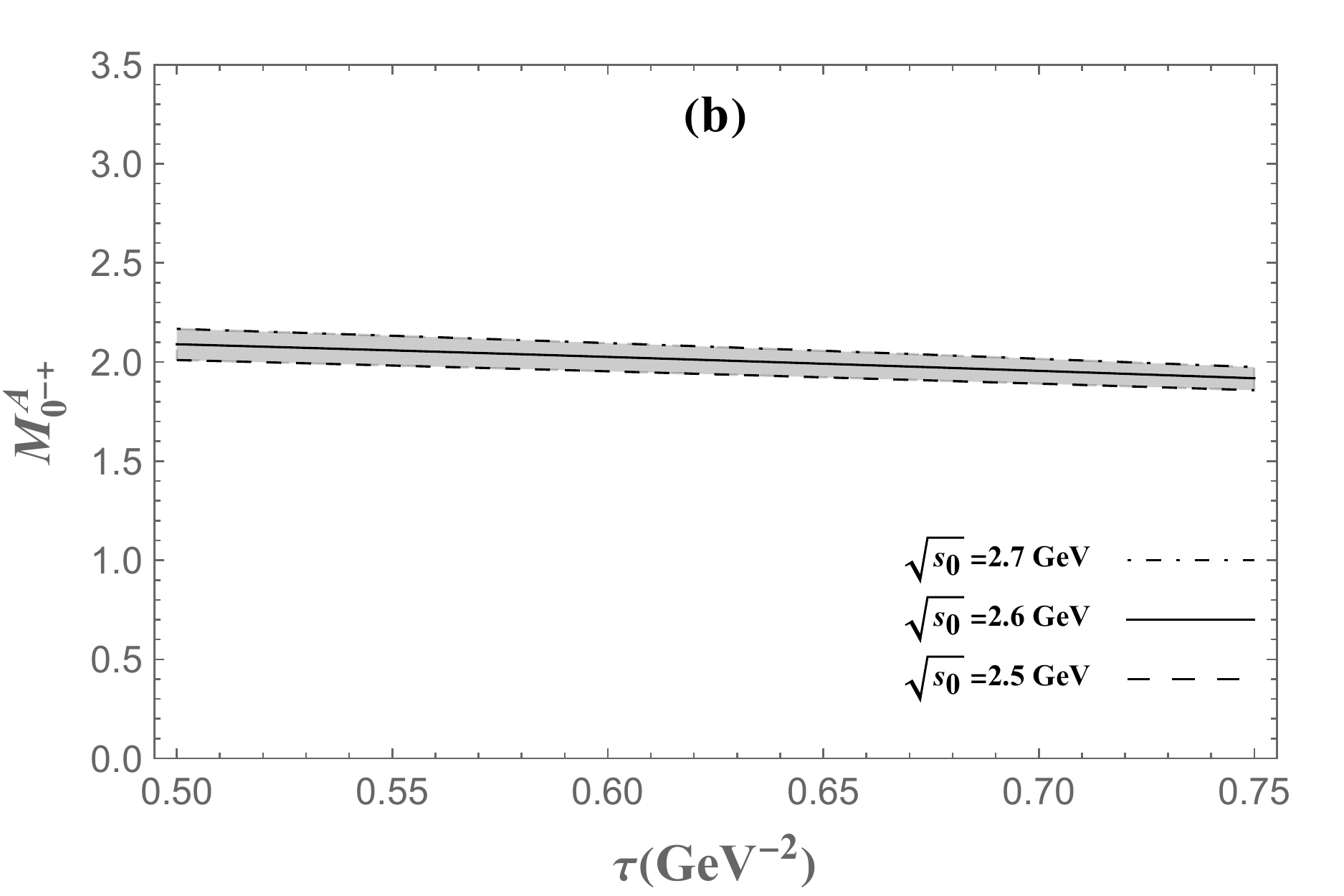}
			\caption{(a) The ratios $R_{0^{-+}}^{A, \text{PC}}$ and $R_{0^{-+}}^{A, \langle \text{G}^3\rangle}$ of case $A$ as functions of the Borel parameter $\tau$ for different values of $\sqrt{s_0}$, where red lines represent $R_{0^{-+}}^{A, \text{PC}}$ and blue lines stand for $R_{0^{-+}}^{A, \langle\text{G}^3\rangle}$. (b) The mass $M_{0^{-+}}^A$ as a function of the Borel parameter $\tau$ for different values of $\sqrt{s_0}$.} \label{fig0-+}
		\end{center}
	\end{figure}

	Through the above preparation, the mass spectrum of the trigluon glueball can be numerically calculated. For the $0^{-+} $ trigluon glueball, the ratios $R^{A, \; \text{PC}}_{0^{-+}} $ and $ R^{A, \; \langle\text{G}^3\rangle}_{0^{-+}}$ as functions of Borel parameter $\tau $ are shown in Fig.~\ref{fig0-+}(a) with different values of $\sqrt{s_0} $, 2.5, 2.6, 2.7 $\text{GeV}$.
	The dependence of the trigluon glueball mass $M_{0^{-+}}^{A} $ on Borel parameter $ \tau $ is shown in Fig.~\ref{fig0-+}(b), in which the upper and lower limits of the Borel parameters $\tau $ are obtained for different values of $\sqrt{s_0} $, as shown in Table.~\ref{table borel}. The situation of case $D$ is exactly the same as case $ A$ because they have common Wilson coefficient $a_1$, $b_1$, $c_1$, and $d_1$ as mentioned above. The mass of the $0^{-+}$ trigluon glueball can then be extracted as follows:
		\begin{eqnarray}\label{mass0}
			\begin{aligned}
					M_{0^{-+}}^{A,\,D} &= \!\!\! & 2.01 \pm 0.14 \, \text{GeV}.
				\end{aligned}
		\end{eqnarray}
	
\begin{figure}[h]
	\begin{center}
		\includegraphics[width=6.8cm]{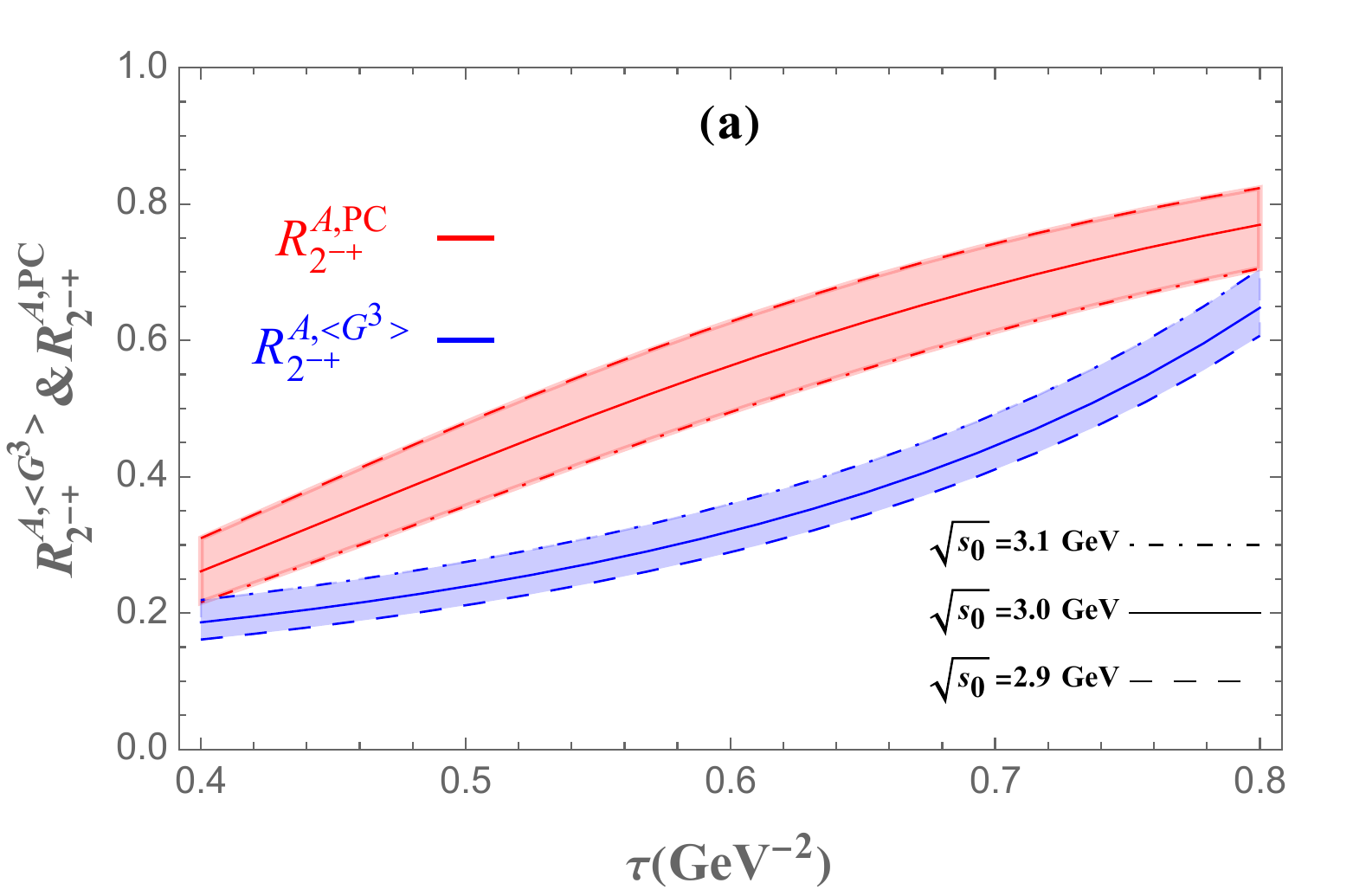}
		\includegraphics[width=6.8cm]{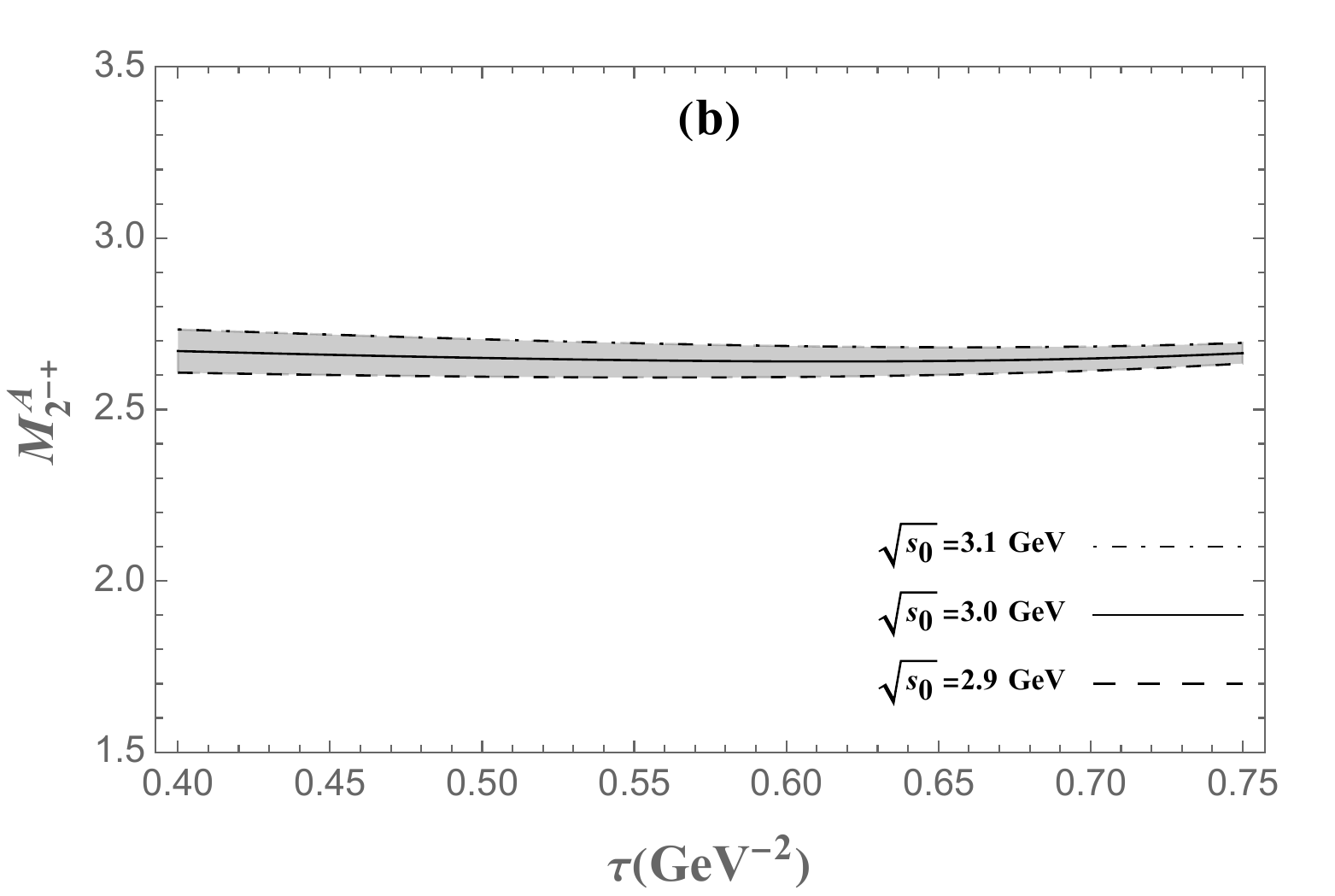}
		\caption{(a) The ratios $R_{2^{-+}}^{A, \text{PC}}$ and $R_{2^{-+}}^{A, \langle\text{G}^3\rangle}$ of case $A$ as functions of the Borel parameter $\tau$ for different values of $\sqrt{s_0}$, where red lines represent $R_{2^{-+}}^{A, \text{PC}}$ and blue lines stand for $R_{2^{-+}}^{A, \langle\text{G}^3\rangle}$. (b) The mass $M_{2^{-+}}^A$ as a function of the Borel parameter $\tau$ for different values of $\sqrt{s_0}$.} \label{fig2-+}
	\end{center}
    \end{figure}

	For the $2^{-+} $ trigluon glueball, we show the corresponding figures of case $ A$ in Fig.~\ref{fig2-+}. The upper and lower limits of the Borel parameters $\tau $ for different values of $\sqrt{s_0} $ are shown in Table~\ref{table borel} and the mass of $2^{-+} $ trigluon glueball can be obtained as follows,
	\begin{eqnarray}\label{mass2}
	\begin{aligned}
		M_{2^{-+}}^{A,\,B,\,D} &= \!\!\! & 2.66 \pm 0.06 \, \text{GeV}.
	\end{aligned}
    \end{eqnarray}

	It should be noted that the errors in masses (\ref{mass0}) and (\ref{mass2}) are mainly determined by the uncertainties of the Borel parameters $ \tau $, continuum threshold $ \sqrt{s_0} $ and condensates given in Eq.~(\ref{input}).
	
	\begin{table}[h]
		\caption{The lower and upper limits of the Borel parameter $\tau$ (GeV$^{-2}$) for $0^{-+}$ and $2^{-+}$ trigluon glueball  with different $\sqrt{s_0}$ (GeV), where $ 0^{-+} $ contains case $ A $ and $ D $, and $ 2^{-+} $ includes case $A$, $ B $, and $D$.}
		\begin{center}
			\begin{tabular}{lllllll}
				\hline
				$ 0^{-+} $\,\,\,\,\,     &               &                                     & \textbf{\,\,\,\,\,} & $ 2^{-+} $\,\,\,\,\,     &               &                                     \\ \hline
				$ \sqrt{s_0} $\,\,\,\,\, & \,\,\,\,\, $\tau_{min} $\,\,\,\,\, & \,\,\,\,\, $\tau_{max} $ \,\,\,\,\, & \,\,\,\,\,          & $ \sqrt{s_0} $\,\,\,\,\, & \,\,\,\,\, $\tau_{min} $\,\,\,\,\, & \,\,\,\,\, $\tau_{max} $ \,\,\,\,\, \\ \hline
				2.5                      & \,\,\,\,\, 0.58 \,\,\,\,\,         & \,\,\,\,\, 0.73 \,\,\,\,\,          & \,\,\,\,\,          & 2.9                      & \,\,\,\,\, 0.53 \,\,\,\,\,         & \,\,\,\,\, 0.70 \,\,\,\,\,          \\
				2.6                      & \,\,\,\,\, 0.54 \,\,\,\,\,         & \,\,\,\,\, 0.74 \,\,\,\,\,          & \,\,\,\,\,          & 3.0                      & \,\,\,\,\, 0.49 \,\,\,\,\,         & \,\,\,\,\, 0.73 \,\,\,\,\,          \\
				2.7                      & \,\,\,\,\, 0.52 \,\,\,\,\,         & \,\,\,\,\, 0.75 \,\,\,\,\,          & \,\,\,\,\,          & 3.1                      & \,\,\,\,\, 0.45 \,\,\,\,\,         & \,\,\,\,\, 0.75 \,\,\,\,\,           \\ \hline
			\end{tabular}
		\end{center}
		\label{table borel}
	\end{table}

    \section{Decay analyses}\label{Decay}
	
    According to preceding calculations, the masses of $ 0^{-+} $ and $ 2^{-+} $ trigluon glueballs in Eqs.~(\ref{mass0}) and (\ref{mass2}) are 2.01 GeV and 2.66 GeV respectively, which are larger than the predictions of corresponding two-gluon glueballs in the literature, as expected\cite{Li:2007ky, Cheng:2008ss, Chanowitz:1980gu, Donoghue:1980hw, Ishikawa:1980xv, Faddeev:2003aw, Close:1996yc, Carlson:1982er}. Our calculation indicates that the mass of $X(2600)$ is close to the $ 2^{-+} $ trigluon glueball. To finally confirm the internal structure of $ X(2600) $, a straightforward procedure is to reconstruct it from its decay products, though its detailed characters still need more researches, especially for states possessing conventional quantum numbers. For comparison, the typical decay channels of two-gluon glueballs given in Refs.~\cite{Li:2007ky, Cheng:2008ss, Chanowitz:1980gu, Donoghue:1980hw, Ishikawa:1980xv, Faddeev:2003aw, Close:1996yc, Carlson:1982er} and trigluon glueballs in this work are shown in Table~\ref{decay channel}. In general, decay modes of three mesons or hadron-antihadron pair for two-gluon glueballs are suppressed, while trigluon glueballs decay to these channels more straightforwardly.
	
    It should also be noted that $X(2600)$ is close in magnitude to baryonium state $\Xi-\bar{\Xi}$ calculated in Ref.~\cite{Wan:2021vny}. The main difference between the baryonium state and the trigluon glueball state is the branching ratio of $N\bar{N} $, $ \Lambda\bar{\Lambda}$, $ \Sigma\bar{\Sigma}$, and $\Xi\bar{\Xi}$ decay mode. For the trigluon glueball, the branching ratios of these four decay modes are close to each other. But for the baryonium state, due to the CKM suppression, the branching ratios of these three $ N\bar{N} $, $ \Lambda\bar{\Lambda}$, and $ \Sigma\bar{\Sigma}$ decay modes are relatively small, as given in Ref.~\cite{Wan:2021vny}, and hence the $\Xi\bar{\Xi}$ channel should be the primary decay mode.

    \begin{table}[ht]
	\caption{Typical decay channels of the two-gluon and trigluon glueball with quantum numbers of $ 0^{-+} $ and $ 2^{-+} $.}
		\begin{tabular}{l|cc}
		\hline
		\multicolumn{1}{c|}{Cases}&   \multicolumn{2}{c}{Possible decay channels}                                                         \\ \hline
		$0^{-+} $ two-gluon glueball $ \rightarrow $ & $\,\,a_0(980)+\pi$                                                                        &       $\,\, \{f_0(500),\,f_0(980)\}+\eta$                                               \\ \hline
		& $\,\, \{f_0(500),\,f_0(980),\,f_0(1370),\,f_0(1500)\}+\eta$                                           &   $\,\,\eta\eta\eta,\,\eta\eta\eta^{\prime},\,\{\eta,\,\eta^{\prime}\}+\pi+\pi$                                 \\
		& $\,\, f_0(500)+f_0(980)+\eta$                                                             & $\,\, \{\omega\omega,\,\rho\rho\}+f_0(500)$                               \\
		$0^{-+} $ trigluon glueball $ \rightarrow $  & $\,\, f_0(500)+f_0(500)+\{\eta,\,\eta^{\prime}\}$                                         & $\,\,N\bar{N} $                                                           \\
		& $\,\, \{f_0(500),\,f_0(980)\}+a_0(980)+\pi$                                               &          \\  \hline
		$2^{-+} $ two-gluon glueball $ \rightarrow $ & $\,\, a_2(1320)+\pi$                                                                      & $\, \, f_0(500)+f_1(1285)$                           \\
		& $\,\, f_2(1270)+\eta$                                                                     &                                                                           \\ \hline
		& $\,\,\eta_2(1645)+f_0(500)$                                                               & $\,\,2 f_1(1285),\,2 a_1(1260),\,2 h_1(1170)$                             \\
		& $\,\, \{f_2(1270),\,f_2^{\prime}(1525)\}+\{\eta,\, \eta^{\prime}\}$                       & $\,\, \rho+\rho+f_0(980)$                                                 \\
		$2^{-+} $ trigluon glueball $ \rightarrow $  & $\,\,a_2(1320)+f_0(500)+\pi$                                                              & $\,\, \{\omega\omega,\,\rho\rho,\,\omega+\phi\}+f_0(500)$                 \\
		& $\,\, \{f_2(1270),\,f_2^{\prime}(1525)\}+f_0(500)+\eta$                                   & $\,\, h_1(1170)+\omega+\eta$                                              \\
		& $\,\,\{f_2(1270),\,f_2^{\prime}(1525)\}+a_0(980)+\pi$                                     & $\,\, \{h_1(1170),\,h_1(1415)\}+\rho+\pi$                                 \\
		& $ \,\,\omega+\phi+\eta,\,  \{\pi\pi,\,\omega\omega,\,\rho\rho\}+\{\eta,\,\eta^{\prime}\}$ & $\,\,N\bar{N},\,\Lambda\bar{\Lambda},\,\Sigma\bar{\Sigma},\,\Xi\bar{\Xi}$ \\ \hline
	\end{tabular}
    \label{decay channel}
    \end{table}

    \section{Conclusions}

    In this work, we investigate the trigluon glueballs with $ J^{PC}=0^{-+} $ and $ 2^{-+} $ which are possible quantum numbers possessed by $X(2600)$ in the framework of QCD sum rules. The numerical results indicate that there are two possible configurations of $0^{-+}$ trigluon glueball states with mass $2.01\pm0.14 $ GeV, and three patterns of $2^{-+}$ trigluon glueball states with mass $2.66\pm0.06 $ GeV.

    We find there are three degenerate $2^{-+}$ trigluon glueballs related to currents (\ref{current-2-+A}), (\ref{current-2-+B}), and (\ref{current-2-+D}), in agreement with the experimental observation of $X(2600)$ in mass within the scope of errors. To disentangle  these states, one needs to find some yet unknown extra conditions, physical observables, to experimentally measure. In case it is realized, the other two disentangled $ 2^{-+} $ trigluon glueballs may be treated as partners of $X(2600)$.

    The possible trigluon glueball decay modes are analyzed and confronted to the corresponding baryonium state $ \Xi-\bar{\Xi}$, by which the internal structure of $X(2600)$ may be decoded with the further experimental verifications, like in BESIII, BELLEII, PANDA, and LHCb experiments.

	\vspace{.5cm} {\bf Acknowledgments} \vspace{.5cm}
	
	This work was supported in part by the National Key Research and Development Program of China under Contracts Nos. 2020YFA0406400; the National Natural Science Foundation of China (NSFC) under the Grants No. 11975236 and No. 11635009.



\begin{thebibliography}{99}
		\bibitem{Gross:1973id}
		D.~J.~Gross and F.~Wilczek,
		Phys. Rev. Lett. \textbf{30}, 1343-1346 (1973).
		
		\bibitem{Politzer:1973fx}
		H.~D.~Politzer,
		Phys. Rev. Lett. \textbf{30}, 1346-1349 (1973).
		
		\bibitem{Wilson:1974sk}
		K.~G.~Wilson,
		Phys. Rev. D \textbf{10}, 2445-2459 (1974).
		
		\bibitem{Meyer:2004gx}
		H.~B.~Meyer,
		[arXiv:hep-lat/0508002 [hep-lat]].
		
		\bibitem{Athenodorou:2020ani}
		A.~Athenodorou and M.~Teper,
		JHEP \textbf{11}, 172 (2020)
		[arXiv:2007.06422 [hep-lat]].
		
		\bibitem{Morningstar:1999rf}
		C.~J.~Morningstar and M.~J.~Peardon,
		Phys. Rev. D \textbf{60}, 034509 (1999)
		[arXiv:hep-lat/9901004 [hep-lat]].
		
		\bibitem{Chen:2005mg}
		Y.~Chen, 
		\textit{et al.},
		Phys. Rev. D \textbf{73}, 014516 (2006)
		[arXiv:hep-lat/0510074 [hep-lat]].
		
		\bibitem{Gregory:2012hu}
		E.~Gregory, 
		\textit{et al.},
		JHEP \textbf{10}, 170 (2012)
		[arXiv:1208.1858 [hep-lat]].
		
		\bibitem{Isgur:1984bm}
		N.~Isgur and J.~E.~Paton,
		Phys. Rev. D \textbf{31}, 2910 (1985).
		
		\bibitem{Faddeev:2003aw}
		L.~Faddeev, A.~J.~Niemi and U.~Wiedner,
		Phys. Rev. D \textbf{70}, 114033 (2004)
		[arXiv:hep-ph/0308240 [hep-ph]].
		
		\bibitem{Jaffe:1975fd}
		R.~L.~Jaffe and K.~Johnson,
		Phys. Lett. B \textbf{60}, 201-204 (1976)
		
		\bibitem{Chodos:1974je}
		A.~Chodos, R.~L.~Jaffe, K.~Johnson, C.~B.~Thorn and V.~F.~Weisskopf,
		Phys. Rev. D \textbf{9}, 3471-3495 (1974).
		
		\bibitem{Donoghue:1980hw}
		J.~F.~Donoghue, K.~Johnson and B.~A.~Li,
		Phys. Lett. B \textbf{99}, 416-420 (1981).
		
		\bibitem{Carlson:1982er}
		C.~E.~Carlson, T.~H.~Hansson and C.~Peterson,
		Phys. Rev. D \textbf{27}, 1556-1564 (1983).

		\bibitem{Szczepaniak:1995cw}
		A.~Szczepaniak, E.~S.~Swanson, C.~R.~Ji and S.~R.~Cotanch,
		Phys. Rev. Lett. \textbf{76}, 2011-2014 (1996)
		[arXiv:hep-ph/9511422 [hep-ph]].
		
		\bibitem{Llanes-Estrada:2005bii}
		F.~J.~Llanes-Estrada, P.~Bicudo and S.~R.~Cotanch,
		Phys. Rev. Lett. \textbf{96}, 081601 (2006)
		[arXiv:hep-ph/0507205 [hep-ph]].
		
		\bibitem{deTeramond:2005su}
		G.~F.~de Teramond and S.~J.~Brodsky,
		Phys. Rev. Lett. \textbf{94}, 201601 (2005)
		[arXiv:hep-th/0501022 [hep-th]].
		
		\bibitem{Brunner:2018wbv}
		F.~Br\"unner, J.~Leutgeb and A.~Rebhan,
		Phys. Lett. B \textbf{788}, 431-435 (2019)
		[arXiv:1807.10164 [hep-ph]].
		
		\bibitem{Zhang:2021itx}
		L.~Zhang, C.~Chen, Y.~Chen and M.~Huang,
		Phys. Rev. D \textbf{105}, 026020 (2022)
		[arXiv:2106.10748 [hep-ph]].
		
        \bibitem{Chen:2015zhh}
        Y.~Chen and M.~Huang,
         Chin. Phys. C \textbf{40}, 123101 (2016)
         [arXiv:1511.07018 [hep-ph]].

        \bibitem{Novikov:1979va}
        V.~A.~Novikov, M.~A.~Shifman, A.~I.~Vainshtein and V.~I.~Zakharov,
        Nucl. Phys. B \textbf{165}, 67-79 (1980).

        \bibitem{Shuryak:1982dp}
        E.~V.~Shuryak,
        Nucl. Phys. B \textbf{203}, 116-139 (1982).

        \bibitem{Narison:1988ts}
        S.~Narison and G.~Veneziano,
        Int. J. Mod. Phys. A \textbf{4}, 2751 (1989).

        \bibitem{Huang:1998wj}
        T.~Huang, H.~Y.~Jin and A.~L.~Zhang,
        Phys. Rev. D \textbf{59}, 034026 (1999)
        [arXiv:hep-ph/9807391 [hep-ph]].

        \bibitem{Forkel:2000fd}
        H.~Forkel,
        Phys. Rev. D \textbf{64}, 034015 (2001)
        [arXiv:hep-ph/0005004 [hep-ph]].

        \bibitem{Bagan:1990sy}
        E.~Bagan and T.~G.~Steele,
        Phys. Lett. B \textbf{243}, 413-420 (1990).

        \bibitem{Chen:2021bck}
        H.~X.~Chen, W.~Chen and S.~L.~Zhu,
        Phys. Rev. D \textbf{104}, 094050 (2021)
        [arXiv:2107.05271 [hep-ph]].

        \bibitem{Chen:2021cjr}
        H.~X.~Chen, W.~Chen and S.~L.~Zhu,
        Phys. Rev. D \textbf{103}, L091503 (2021)
        [arXiv:2103.17201 [hep-ph]].

		\bibitem{Shifman:1978by}
		M.~A.~Shifman, A.~I.~Vainshtein and V.~I.~Zakharov,
		Nucl. Phys. B \textbf{147}, 448-518 (1979).
		
		\bibitem{Hao:2005hu}
		G.~Hao, C.~F.~Qiao and A.~L.~Zhang,
		Phys. Lett. B \textbf{642}, 53-61 (2006)
		[arXiv:hep-ph/0512214 [hep-ph]].
		
		\bibitem{Qiao:2014vva}
		C.~F.~Qiao and L.~Tang,
		Phys. Rev. Lett. \textbf{113}, 221601 (2014)
		[arXiv:1408.3995 [hep-ph]].
		
		\bibitem{Tang:2015twt}
		L.~Tang and C.~F.~Qiao,
		Nucl. Phys. B \textbf{904}, 282-296 (2016)
		[arXiv:1509.00305 [hep-ph]].
		
		\bibitem{Narison:2021xhc}
		S.~Narison,
		Nucl. Phys. A \textbf{1017}, 122337 (2022)
		[arXiv:2108.13089 [hep-ph]].
		
		\bibitem{BES:2004twe}
		M.~Ablikim \textit{et al.} [BES],
		Phys. Lett. B \textbf{607}, 243-253 (2005)
		[arXiv:hep-ex/0411001 [hep-ex]].
		
		\bibitem{Gray:1983cw}
		L.~Gray, T.~Kalogeropoulos, A.~Nandy, J.~Roy and S.~Zenone,
		Phys. Rev. D \textbf{27}, 307-310 (1983).
		
		\bibitem{Serpukhov-Brussels-AnnecyLAPP:1983xdr}
		F.~G.~Binon \textit{et al.} [Serpukhov-Brussels-Annecy(LAPP)],
		Nuovo Cim. A \textbf{78}, 313 (1983)
		CERN-EP/83-97.
		
		\bibitem{Burke:1982am}
		D.~L.~Burke, 
		\textit{et al.},
		Phys. Rev. Lett. \textbf{49}, 632 (1982).
		
		\bibitem{Etkin:1982se}
		A.~Etkin, 
		\textit{et al.},
		Phys. Rev. D \textbf{25}, 2446 (1982).
		
		\bibitem{Edwards:1981ex}
		C.~Edwards, 
		\textit{et al.},
		Phys. Rev. Lett. \textbf{48}, 458 (1982).
		
		\bibitem{Cheng:2008ss}
		H.~Y.~Cheng, H.~n.~Li and K.~F.~Liu,
		Phys. Rev. D \textbf{79}, 014024 (2009)
		[arXiv:0811.2577 [hep-ph]].
		
		\bibitem{Li:2007ky}
		G.~Li, Q.~Zhao and C.~H.~Chang,
		J. Phys. G \textbf{35}, 055002 (2008)
		[arXiv:hep-ph/0701020 [hep-ph]].
		
		\bibitem{Chanowitz:1980gu}
		M.~S.~Chanowitz,
		Phys. Rev. Lett. \textbf{46}, 981 (1981).
		
		\bibitem{Ishikawa:1980xv}
		K.~Ishikawa,
		Phys. Rev. Lett. \textbf{46}, 978 (1981)
		
		\bibitem{Close:1996yc}
		F.~E.~Close, G.~R.~Farrar and Z.~p.~Li,
		Phys. Rev. D \textbf{55}, 5749-5766 (1997)
		[arXiv:hep-ph/9610280 [hep-ph]].
		
		\bibitem{BESIII:2011nqb}
		M.~Ablikim \textit{et al.} [BESIII],
		Phys. Rev. Lett. \textbf{107}, 182001 (2011)
		[arXiv:1107.1806 [hep-ex]].
		
		\bibitem{Wan:2019fuk}
		B.~D.~Wan and C.~F.~Qiao,
		Chin. Phys. C \textbf{44}, 093105 (2020)
		[arXiv:1904.02067 [hep-ph]].
		
		\bibitem{Chen:2022asf}
		H.~X.~Chen, W.~Chen, X.~Liu, Y.~R.~Liu and S.~L.~Zhu,
		[arXiv:2204.02649 [hep-ph]].
		
		\bibitem{BES:2005ega}
		M.~Ablikim \textit{et al.} [BES],
		Phys. Rev. Lett. \textbf{95}, 262001 (2005)
		[arXiv:hep-ex/0508025 [hep-ex]].
		
		\bibitem{BESIII:2016fbr}
		M.~Ablikim \textit{et al.} [BESIII],
		Phys. Rev. Lett. \textbf{117}, 042002 (2016)
		[arXiv:1603.09653 [hep-ex]].
		
		\bibitem{BESIII:2021xoh}
		M.~Ablikim \textit{et al.} [BESIII],
		Phys. Rev. Lett. \textbf{129}, 2, 022002 (2022)
		[arXiv:2112.14369 [hep-ex]].
		
		\bibitem{BESIII:2020kpr}
		M.~Ablikim \textit{et al.} [BESIII],
		Phys. Rev. D \textbf{103}, 072007 (2021)
		[arXiv:2012.07360 [hep-ex]].
		
		\bibitem{BESIII:2019sfz}
		M.~Ablikim \textit{et al.} [BESIII],
		Phys. Rev. D \textbf{99}, 071101 (2019)
		[arXiv:1902.04862 [hep-ex]].
		
		\bibitem{BESIII:2017hup}
		M.~Ablikim \textit{et al.} [BESIII],
		Phys. Rev. D \textbf{96}, 112012 (2017)
		[arXiv:1709.00018 [hep-ex]].
		
		\bibitem{BESIII:2010gmv}
		M.~Ablikim \textit{et al.} [BESIII],
		Phys. Rev. Lett. \textbf{106}, 072002 (2011)
		[arXiv:1012.3510 [hep-ex]].
		
		\bibitem{Amsler:2004ps}
		C.~Amsler and N.~A.~Tornqvist,
		Phys. Rept. \textbf{389}, 61-117 (2004).
		
		\bibitem{Eshraim:2012jv}
		W.~I.~Eshraim, S.~Janowski, F.~Giacosa and D.~H.~Rischke,
		Phys. Rev. D \textbf{87}, 054036 (2013)
		[arXiv:1208.6474 [hep-ph]].
		
		\bibitem{BESIIICollaboration:2022qjc}
		M.~Ablikim \textit{et al.} [(BESIII Collaboration)* and BESIII],
		Phys. Rev. Lett. \textbf{129}, 4, 042001 (2022)
		[arXiv:2201.10796 [hep-ex]].
		
		\bibitem{Qiao:2013dda}
		C.~F.~Qiao and L.~Tang,
		Eur. Phys. J. C \textbf{74}, 2810 (2014)
		[arXiv:1308.3439 [hep-ph]].
		
		\bibitem{Chen:2011qu}
		W.~Chen, Z.~X.~Cai and S.~L.~Zhu,
		Nucl. Phys. B \textbf{887}, 201-215 (2014)
		[arXiv:1107.4949 [hep-ph]].

		\bibitem{Ioffe:2005ym}
		B.~L.~Ioffe,
		Prog. Part. Nucl. Phys. \textbf{56}, 232-277 (2006)
		[arXiv:hep-ph/0502148 [hep-ph]].
		
		\bibitem{Reinders:1984sr}
		L.~J.~Reinders, H.~Rubinstein and S.~Yazaki,
		Phys. Rept. \textbf{127}, 1 (1985).
		
		\bibitem{Colangelo:2000dp}
		P.~Colangelo and A.~Khodjamirian,
		[arXiv:hep-ph/0010175 [hep-ph]].

		\bibitem{Wan:2021vny}
		B.~D.~Wan, S.~Q.~Zhang and C.~F.~Qiao,
		Phys. Rev. D \textbf{105}, 014016 (2022)
		[arXiv:2109.07130 [hep-ph]].
		
		\bibitem{Wan:2019ake}
		B.~D.~Wan, L.~Tang and C.~F.~Qiao,
		Eur. Phys. J. C \textbf{80}, 2, 121 (2020)
		[arXiv:1912.12046 [hep-ph]].
		
		\bibitem{Wang:2016gxp}
		Z.~G.~Wang,
		Eur. Phys. J. C \textbf{77}, 78 (2017)
		[arXiv:1606.05872 [hep-ph]].
		
		\bibitem{Tang:2021zti}
		C.~M.~Tang, Y.~C.~Zhao and L.~Tang,
		Phys. Rev. D \textbf{105}, 11, 114004 (2022)
		[arXiv:2111.07328 [hep-ph]].
		
		\bibitem{Wan:2022xkx}
		B.~D.~Wan, S.~Q.~Zhang and C.~F.~Qiao,
		[arXiv:2203.14014 [hep-ph]].
		
		\bibitem{Xin:2021wcr}
		Q.~Xin and Z.~G.~Wang,
		Eur. Phys. J. A \textbf{58}, 6, 110 (2022)
		[arXiv:2108.12597 [hep-ph]].
		
		\bibitem{Wang:2021qus}
		Z.~G.~Wang,
		Nucl. Phys. B \textbf{973}, 115592 (2021)
		[arXiv:2108.05759 [hep-ph]].
		
		\bibitem{Wan:2020fsk}
		B.~D.~Wan and C.~F.~Qiao,
		Phys. Lett. B \textbf{817}, 136339 (2021)
		[arXiv:2012.00454 [hep-ph]].
		
		\bibitem{Wan:2020oxt}
		B.~D.~Wan and C.~F.~Qiao,
		Nucl. Phys. B \textbf{968}, 115450 (2021)
		[arXiv:2011.08747 [hep-ph]].
		
		\bibitem{Yang:2020wkh}
		B.~C.~Yang, L.~Tang and C.~F.~Qiao,
		Eur. Phys. J. C \textbf{81}, 4, 324 (2021)
		[arXiv:2012.04463 [hep-ph]].
	
    \end{thebibliography}
\end{document}